\documentclass[12pt]{amsart}

\usepackage{amsmath,amssymb,cite,enumerate,graphicx,hyperref}

\hypersetup{pdftex,colorlinks=true,allcolors=blue}
\usepackage[all]{hypcap}

\usepackage[foot]{amsaddr}
\usepackage[sc]{mathpazo}

\newcommand\ack{\subsection*{Acknowledgment}}

\DeclareMathAlphabet\mathsfbi{T1}{phv}{b}{it}

\numberwithin{equation}{section}

\newcommand\BV{\boldsymbol} 
\newcommand\BM{\mathsfbi} 
\newcommand\dif{\,\mathrm{d}}
\newcommand\deriv[2]{\frac{\mathrm{d} #1}{\mathrm{d} #2}}
\newcommand\parderiv[2]{\frac{\partial #1}{\partial #2}}

\newcommand\cF{\mathcal F}

\newcommand\Hside{\,\Theta}

\newcommand\Mach{M}

\newcommand\CMach{\widetilde\Mach}

\newcommand\MolarMass{\mathcal M}
\newcommand\GasConstant{\mathcal R}

\begin{document}

\author[Rafail V. Abramov]{Rafail V. Abramov}

\address{Department of Mathematics, Statistics and Computer Science,
University of Illinois at Chicago, 851 S. Morgan st., Chicago, IL 60607}

\email{abramov@uic.edu}

\title[The effect of the Enskog collision terms on the steady shock
  structure]{The effect of the Enskog collision terms on the steady
  shock structure in a hard sphere gas}

\begin{abstract}
In this work we study the effect of the Enskog collision terms on the
steady shock transitions in the supersonic flow of a hard sphere
gas. We start by examining one-dimensional, nonlinear, nondispersive
planar wave solutions of the Enskog--Navier--Stokes equations, which
move in a fixed direction at a constant speed. By further equating the
speed of the reference frame with the speed of such a wave, we reduce
the Enskog--Navier--Stokes equations into a more simple system of two
ordinary differential equations, whose solutions depend on a single
scalar spatial variable. We then observe that this system has two
fixed points, which are taken to be the states of the gas before and
after the shock, and compute the corresponding shock transition in the
form of the heteroclinic orbit connecting these two states. We find
that the Enskog correction affects both the difference between the
fixed points, and the thickness of the transition. In particular, for
a given state of the gas before the shock transition, the difference
between the fixed points is reduced, while the shock thickness is
increased, with the relative impact on the properties of transition
being more prominent at low Mach numbers. We also compute the speed of
sound in the Enskog--Navier--Stokes equations, and find that, for the
same thermodynamic state, it is somewhat faster than that in the
conventional Navier--Stokes equations, with an additional dependence
on the density of the gas.
\end{abstract}

\maketitle

\section{Introduction}

Conventionally, the hard sphere gas is modeled by the famous Boltzmann
equation \cite{Bol,Cer,Cer2,CerIllPul}. The corresponding fluid
dynamics approximation is given by the Navier--Stokes equations
\cite{Bat,Gols}. This conventional theory is used in many engineering
applications, and is a good approximation for gases at normal
conditions (that is, not too rarefied, and not too dense).

However, in a number of works
\cite{Brenner,Brenner2,Brenner3,DadRee,Durst,Durst2,GreRee} it was
noted that, in certain situations, the conventional model, offered via
the Boltzmann equation and the Navier--Stokes equations, failed to
capture some features of the flow of a real gas. In particular, in
\cite{GreRee} it was observed that the shock transition thickness in a
real gas was not captured accurately by the conventional
Navier--Stokes equations; the latter produced a much thinner shock
transition than what was actually observed. It was speculated in
\cite{Brenner,Brenner2,Brenner3,DadRee,Durst,Durst2,Kli,Sher} that an
\textsl{ad hoc} diffusive term could mitigate some of the
discrepancies between the observed and modeled behavior. In
\cite{Abr13,AbrOtt} we studied a diffusive higher-order approximation
to the gas dynamics, based on the Grad moment truncation
\cite{Gra,StruTor}.

In \cite{Abr17}, we proposed a new model of the hard sphere gas, based
on random interactions. This model consists of a random jump process
\cite{App,Fel2,DalVer,Papa}, whose infinitesimal generator \cite{Cou}
allows a straightforward construction of the forward Kolmogorov
equation for the probability density of states of the full
multi-sphere system. We found a physically realistic family of the
steady states, and, with its help, constructed the forward equation
for the marginal statistical distribution of a single sphere. This
equation turned out to be the Enskog equation \cite{Ens}, originally
obtained by Enskog as an \textsl{ad hoc} model of a dense gas. In
\cite{Fre}, it was found that the Enskog equation produced accurate
shock transitions for a wide range of regimes.

In addition to the new random hard sphere model, in \cite{Abr17} we
proposed a new hydrodynamic limit, where the density of each sphere of
gas is kept constant while its diameter becomes small. This limit is
different from the conventional Boltzmann--Grad limit \cite{Gra}, and
appears to be more realistic for noble gases. In the constant-density
limit, we found that the resulting Enskog--Euler \cite{Lac} and
Enskog--Navier--Stokes equations acquire additional corrections, both
in the advective and viscous terms. In the present work, we study the
shock transitions produced by the Enskog--Navier--Stokes equations
\cite{Abr17}, and compare them with those produced by the conventional
Navier--Stokes equations \cite{Bat,Gols}.

\subsection{The Enskog--Navier--Stokes equations}

In \cite{Abr17}, we derived the Enskog--Navier--Stokes equations for
the hard sphere gas. What follows is a rough sketch of the procedure
of derivation:
\begin{enumerate}
\item We found that the conventional derivation \cite{CerIllPul} of
  the Boltzmann equation had logical inconsistencies. In particular,
  we observed that solutions of the Boltzmann equation violated
  assumptions, under which the latter was derived (see Section 3 of
  \cite{Abr17} for more details).
\item However, the Boltzmann equation is known to be a very accurate
  model of a gas, which means that the equation itself is valid, even
  though its conventional derivation is flawed. To this end, we
  introduced a mathematically consistent model of the hard sphere gas,
  which uses a random jump process with conditional intensity
  \cite{App,Fel,DalVer,Papa} at its foundation (see Section 4 of
  \cite{Abr17} for more details).
\item The Bogoliubov--Born--Green--Kirkwood--Yvon formalism
  \cite{Bog,BorGre,Kir}, applied to the random hard sphere model in
  \cite{Abr17}, yields a variant of Enskog equation \cite{HirCurBir}
  for $K$ spheres in the form
\begin{multline}
\label{eq:enskog}
\parderiv ft+\BV v\cdot\parderiv f{\BV x}=(K-1)\sigma^2 R(\sigma)
\int\BV n\cdot(\BV w-\BV v)\Hside\big(\BV n\cdot (\BV w-\BV
v)\big)\\\left[f(\BV x,\BV v')f(\BV x+\sigma\BV n,\BV w') -f(\BV x,\BV
  v)f(\BV x-\sigma\BV n,\BV w)\right]\dif\BV n\dif\BV w.
\end{multline}
Above, $f(t,\BV x,\BV v)$ is the probability density of distribution
of a single molecule over the velocities and coordinates (same as in
the Boltzmann equation), $\sigma$ is the diameter of a sphere, $\BV n$
is a vector on a unit sphere, while $\Hside(x)$ is the usual Heaviside
function. In the collision integral, the recedent velocities $\BV v'$
and $\BV w'$ are given via the usual deflection formulas
\cite{Cer,Cer2,CerIllPul}
\begin{equation}
  \BV v'=\BV v+\big((\BV w-\BV v)\cdot\BV n\big)\BV n,\qquad \BV
  w'=\BV w+\big((\BV v-\BV w)\cdot\BV n\big)\BV n.
\end{equation}
The main difference between the collision integrals of the
conventional Enskog equation (see, e.g. Eqn. (9.3-6) in
\cite{HirCurBir}, or Eqn. (1) in either \cite{Fre2} or \cite{Fre}) and
those of the random hard sphere Enskog equation in \eqref{eq:enskog}
and \cite{Abr17} is that the collision integral of the former contains
an empirical factor obtained via the virial expansion of the equation
of state of the gas (see Eqn. (9.3-2) in \cite{HirCurBir}), while the
collision integral of \eqref{eq:enskog} is prepended by the spatial
correlation function $R(\sigma)$. This correlation function is the
result of the systematic closure of the joint two-particle
distribution via two single-particle distributions, and the exact
expression for $R(\sigma)$ is given by the Eqn. (146) in
\cite{Abr17}. As a result, the gas is not required to be {\em dilute}
in the Enskog equation \eqref{eq:enskog}, that is, the distances
between the spheres are not required to be much larger than their
diameters, and the spheres themselves are no longer presumed to be
distributed statistically independently. See Section 6 in \cite{Abr17}
for more details on the derivation of the Enskog equation in
\eqref{eq:enskog}.
\item In order to derive the Enskog--Navier--Stokes equations, the
  probability density in \eqref{eq:enskog} must be weighted by the
  total mass of the system of $K$ spheres. To account for that, in
  \cite{Abr17} we endowed each sphere with mass density $\rho_{sp}$,
  and introduced the mass-weighted probability distribution
  \begin{equation}
    g(t,\BV x,\BV v)=Kmf(t,\BV x,\BV v),\qquad m=\frac 16\pi\rho_{sp}
    \sigma^3.
  \end{equation}
  Substituting the expression for $g$ into \eqref{eq:enskog}, and
  taking the number of spheres $K\to\infty$, in \cite{Abr17} we
  arrived at the mass-weighted Enskog equation:
\begin{multline}
\label{eq:mass_enskog}
\parderiv gt+\BV v\cdot\parderiv g{\BV x}=\frac{6R(\sigma)}{\pi
  \rho_{sp}\sigma}\int\BV n\cdot(\BV w-\BV v)\Hside\big(\BV n\cdot
(\BV w-\BV v)\big)\\\left[g(\BV x,\BV v')g(\BV x+\sigma\BV n,\BV w')
  -g(\BV x,\BV v)g(\BV x-\sigma\BV n,\BV w)\right]\dif\BV n\dif\BV w.
\end{multline}
\item Recall that the conventional Navier--Stokes equations are
  obtained from the Boltzmann equation in the Boltzmann--Grad
  hydrodynamic limit \cite{Gra}. In this limit, the diameter of the
  sphere is taken to zero, while the mass-weighted collision
  cross-section is kept constant (that is, $\sigma\to 0$,
  $\sigma^2/m=$ const).  However, this leads to $\rho_{sp}\to\infty$,
  which is physically unrealistic. As we demonstrated in Fig. 1 of
  \cite{Abr17}, the atoms of noble gases have roughly constant mass
  density, and therefore, the constant-density hydrodynamic limit
  ($\sigma\to 0$, $\rho_{sp}=$ const) is physically more realistic.
\end{enumerate}
Expanding the mass-weighted Enskog equation in \eqref{eq:mass_enskog}
in powers of $\sigma$, taking the constant sphere density hydrodynamic
limit (instead of the conventional Boltzmann--Grad limit), and
repeating the same derivation steps as are used to arrive from the
Boltzmann equation to the conventional Navier--Stokes equations
\cite{Gra}, in \cite{Abr17} we arrived at the following
Enskog--Navier--Stokes equations in the leading order of the
expansion:
\begin{subequations}
\label{eq:enskog_navier_stokes}
\begin{equation}
 \parderiv\rho t+\parderiv{}{\BV x}\cdot(\rho\BV u)=0,
\end{equation}
\begin{equation}
\parderiv{(\rho\BV u)}t+\parderiv{}{\BV x}\cdot\big(\rho\BV u\BV u^T
+p\BM I\big)=\parderiv{}{\BV x}\cdot\bigg\{\mu\bigg[(1+a_1)\bigg(
  \parderiv{ \BV u}{\BV x}+\parderiv{\BV u}{\BV x}^T\bigg)-\frac 23
  (1+a_2)\bigg( \parderiv{}{\BV x}\cdot\BV u\bigg)\BM I\bigg]\bigg\},
\end{equation}
\begin{multline}
\parderiv{(\rho\epsilon)}t+\parderiv{}{\BV x}\cdot\big((\rho\epsilon
+p)\BV u\big)=\frac{15}4\parderiv{}{\BV x}\cdot\bigg(\mu(1+a_3)
\parderiv\theta{\BV x}\bigg)+\\+\parderiv{}{\BV x}\cdot\bigg\{\mu
\bigg[(1+a_1) \bigg(\parderiv{\BV u}{\BV x}+\parderiv{\BV u}{\BV x
}^T\bigg)-\frac 23(1+a_2)\bigg(\parderiv{}{\BV x}\cdot\BV u\bigg)\BM
I\bigg]\BV u\bigg\}.
\end{multline}
\end{subequations}
Above, $\rho$, $\BV u$ and $\epsilon$ are the density, velocity, and
kinetic energy of the gas, given via the standard velocity moments
\begin{equation}
\label{eq:rho_u_e}
\rho=\int g\dif\BV v,\qquad\rho\BV u=\int\BV vg\dif\BV v,\qquad
\rho\epsilon=\frac 12\int\|\BV v\|^2g\dif\BV v,
\end{equation}
while $\theta$ and $p$ are, respectively, the kinetic temperature and
pressure:
\begin{equation}
\label{eq:theta_p}
\theta=\frac 1{3\rho}\int\|\BV v-\BV u\|^2g\dif\BV v=\frac
13(2\epsilon-\|\BV u\|^2),\qquad p=\rho\theta
\left(1+\frac{4\rho}{\rho_{sp}}\right).
\end{equation}
Note that, in \cite{Abr17}, the equations in
\eqref{eq:enskog_navier_stokes} are formulated entirely in terms of
$\rho$, $\BV u$, $\theta$ and $\epsilon$; here we introduce the
pressure variable $p$ in \eqref{eq:theta_p} primarily for convenience
-- although, the expression for $p$ in \eqref{eq:theta_p} can also be
interpreted as a simplified van der Waals ``equation of state'' with
parameters $a=0,b=4/\rho_{sp}$.

Observe that setting $\rho/\rho_{sp}=0$ (which corresponds to the
Boltzmann--Grad limit with $\rho_{sp}\to\infty$) in the definition of
the pressure $p$ in \eqref{eq:theta_p} reverts the latter back to its
usual definition, that is, the product of the density with kinetic
temperature \cite{Gra,Gols}. The conventional temperature $T$ is
related to $\theta$ via
\begin{equation}
\label{eq:temperature}
T=\frac\MolarMass\GasConstant\theta,
\end{equation}
where $\GasConstant=8.314$ kg m$^2$/(mol K s$^2$) is the universal gas
constant, $\MolarMass$ is the molar mass of the gas. The viscosity
$\mu$, and the correction coefficients $a_1$, $a_2$ and $a_3$ are
given via
\begin{subequations}
\label{eq:viscosity}
\begin{equation}
\mu=\frac{5\sqrt\pi\rho_{sp}\sigma}{96}\sqrt\theta,
\qquad a_1=\frac{16\rho}{5\rho_{sp}}\left(1+\frac{4\rho}{5\rho_{sp}}
\left(1+\frac{12}\pi\right)\right),
\end{equation}
\begin{equation}
 a_2=\frac{16
  \rho}{5\rho_{sp}}\left(1+\frac{4 \rho}{5\rho_{sp}}
\left(1-\frac{18}\pi\right)\right),
\qquad
a_3=\frac{24\rho}{5\rho_{sp}}\left(1+ \frac{2\rho}{15\rho_{sp}}
\left(9+\frac{32}\pi\right)\right).
\end{equation}
\end{subequations}
Again, observe that setting $\rho/\rho_{sp}=0$, which corresponds to
the Boltzmann--Grad limit with $\rho_{sp}\to\infty$, sets
$a_1=a_2=a_3=0$, and reverts the equations in
\eqref{eq:enskog_navier_stokes} back to the conventional
Navier--Stokes equations \cite{Bat,Gols} for a monatomic gas.

The main advantage of the obtained Enskog--Navier--Stokes equations in
\eqref{eq:enskog_navier_stokes}--\eqref{eq:viscosity} is that the gas
is no longer required to be dilute, that is, average distances between
the spheres are no longer required to be much larger than the
diameters of the spheres themselves. As a result, the ratio
$\rho/\rho_{sp}$ is no longer presumed to be zero, and is present in
the equations as a correction term. For more details on the derivation
of \eqref{eq:enskog_navier_stokes}--\eqref{eq:viscosity}, see Section
7 of \cite{Abr17}.

In this work, we study the properties of the shock wave transitions
for the Enskog--Navier--Stokes equations in
\eqref{eq:enskog_navier_stokes}.  The paper is organized as
follows. In Section \ref{sec:sound} we derive the acoustic equations
and determine the corrected speed of sound for the
Enskog--Navier--Stokes equations. We find that, due to the Enskog
correction, the actual speed of sound in a hard sphere gas is somewhat
faster than predicted by the conventional Navier--Stokes equations. In
Section \ref{sec:steady} we consider the steady one-dimensional
scenario, and reduce the Enskog--Navier--Stokes dynamics to the system
of two ordinary differential equations. In Section \ref{sec:fixed} we
study the fixed points of the system of ordinary differential
equations, which correspond to the states of the gas before and after
the shock. We find that the Enskog correction tends to reduce the
difference between the fixed points, and that, in particular, for a
given state before the shock, the Enskog-corrected dynamics produce
the post-shock state which has lower density and temperature, and
higher velocity. In Section \ref{sec:transitions} we examine the
heteroclinic orbit \cite{GucHol} which connects the fixed
points, and which comprises the shock transition. We find that the
Enskog-corrected model tends to increase the thickness of the shock
transition in comparison with the conventional Navier--Stokes
equations, particularly at transonic Mach numbers.

\section{The acoustic equations and the speed of sound}
\label{sec:sound}

Here we compute the corrected speed of sound and the formula for the
corresponding Mach number for the Enskog--Navier--Stokes equations in
\eqref{eq:enskog_navier_stokes}. Assuming homogeneity in $y$- and
$z$-directions, we write the Enskog--Navier--Stokes equations as
\begin{subequations}
\label{eq:enskog_navier_stokes_1D}
\begin{equation}
\parderiv\rho t+\parderiv{(\rho u)}x=0,
\end{equation}
\begin{equation}
\parderiv{(\rho u)}t+\parderiv{}x \left(\rho u^2+p\right)=\frac
43\parderiv{}x \left(\mu\left(1+\frac 32a_1-\frac
12a_2\right)\parderiv ux\right),
\end{equation}
\begin{equation}
\parderiv{(\rho\epsilon)}t+\parderiv{}x\big((\rho\epsilon +p)u\big)=
\frac{15}4\parderiv{}x\left(\mu(1+a_3)\parderiv\theta x\right)+\frac
43\parderiv{}x\left(\mu\left(1+\frac 32a_1-\frac 12a_2\right) u
\parderiv ux\right).
\end{equation}
\end{subequations}
Expressing $\epsilon$ and $p$ via \eqref{eq:theta_p}, and following
the same procedure as in \cite{Gols}, we assume $\rho=\bar\rho+\rho'$,
$u=u'$, $\theta=\bar\theta+\theta'$, where the overlined terms denote
the uniform background state, while the primed notations denote small
fluctuations. Then, linearizing \eqref{eq:enskog_navier_stokes_1D}
near the background state and omitting the viscous terms, we arrive at
\begin{subequations}
\begin{equation}
\label{eq:enskog_navier_stokes_1D_linearized_mass}
\parderiv{\rho'}t+\bar\rho\parderiv{u'}x=0,
\end{equation}
\begin{equation}
\label{eq:enskog_navier_stokes_1D_linearized_momentum}
\bar\rho\parderiv{u'}t+\left(1+\frac{4\bar\rho}{\rho_{sp}} \right)
\bar\rho\parderiv{\theta'}x +\left(1+\frac{8\bar\rho}{\rho_{sp}}
\right)\bar\theta\parderiv{\rho'}x=0,
\end{equation}
\begin{equation}
\label{eq:enskog_navier_stokes_1D_linearized_energy}
\frac 32\bar\rho\parderiv{\theta'}t+\frac 32\bar\theta \parderiv{
  \rho'}t+\left(\frac 52 +\frac{4\bar\rho}{ \rho_{sp}}\right)
\bar\rho\bar\theta\parderiv{u'}x =0.
\end{equation}
\end{subequations}
The idea now is to balance the time-derivatives in
\eqref{eq:enskog_navier_stokes_1D_linearized_energy} above in the same
proportions as the corresponding spatial derivatives in
\eqref{eq:enskog_navier_stokes_1D_linearized_momentum}, with the help
of \eqref{eq:enskog_navier_stokes_1D_linearized_mass}. First, we
rescale \eqref{eq:enskog_navier_stokes_1D_linearized_energy} as
\begin{equation}
\left(1+\frac{4\bar\rho}{\rho_{sp}}\right)\bar\rho\parderiv{\theta'}t
+\left(1+\frac{4\bar\rho}{\rho_{sp}}\right)\bar\theta\parderiv{\rho'}
t=-\frac 23\left(1+\frac{4\bar\rho}{\rho_{sp}} \right)\left(\frac 52
+\frac{4\bar\rho}{\rho_{sp}}\right) \bar\rho\bar\theta\parderiv{u'}x.
\end{equation}
Second, we rescale \eqref{eq:enskog_navier_stokes_1D_linearized_mass}
as
\begin{equation}
\frac{4\bar\rho}{\rho_{sp}}\bar\theta\parderiv{\rho'}t=-\frac{4\bar
  \rho}{\rho_{sp}}\bar\rho\bar\theta\parderiv{u'}x.
\end{equation}
Adding the results together, we arrive at
\begin{equation}
\parderiv{}t\left(\left(1+\frac{4\bar\rho}{\rho_{sp}}\right)\bar\rho
  \theta'+\left(1+\frac{8\bar\rho}{\rho_{sp}}\right)\bar\theta\rho'
  \right)=-\frac 53\left(1+\frac{8\bar\rho}{\rho_{sp}}+\frac{32
  \bar\rho^2}{5\rho_{sp}^2}\right)\bar\rho\bar\theta\parderiv{u'}x.
\end{equation}
Differentiating in $t$ and dividing by the product
$\bar\rho\bar\theta$, we have
\begin{equation}
\parderiv{^2}{t^2}\left(\left(1+\frac{8\bar\rho}{\rho_{sp}}\right)
\frac{\rho'}{\bar\rho}+\left(1+\frac{4\bar\rho}{\rho_{sp}}\right)
\frac{\theta'}{\bar\theta}\right)=-\frac 53\left(1+\frac{8\bar\rho}{
  \rho_{sp}}+\frac{32\bar\rho^2}{5\rho_{sp}^2}\right)\parderiv{^2u'}{
  t\partial x}.
\end{equation}
Now, we differentiate
\eqref{eq:enskog_navier_stokes_1D_linearized_momentum} in $x$,
arriving at
\begin{equation}
\parderiv{^2}{x^2}\left(\left(1+\frac{8\bar\rho}{\rho_{sp}}\right)\frac{
    \rho'}{\bar\rho}+\left(1+\frac{4\bar\rho}{\rho_{sp}} \right)
  \frac{\theta'}{\bar\theta}\right)=-\frac 1{\bar\theta}
\parderiv{^2u'}{t\partial x}.
\end{equation}
Equating the mixed derivatives in the last two equations, we obtain
\begin{equation}
\left[\parderiv{^2}{t^2}-\frac 53\bar\theta\left(1+\frac{8\bar\rho}{
    \rho_{sp}}+\frac{32\bar\rho^2}{5\rho_{sp}^2}\right)\parderiv{^2}{
    x^2}\right]\left(\left(1+\frac{8\bar\rho}{\rho_{sp}}\right)\frac{
  \rho'}{\bar\rho}+\left(1+\frac{4\bar\rho}{\rho_{sp}}\right)\frac{
  \theta'}{\bar\theta}\right)=0.
\end{equation}
We arrived at the Enskog-corrected version of the acoustic equation
from \cite{Gols}. The speed of sound $c$ is, obviously, given via the
square root of the coefficient in front of the spatial second
derivative:
\begin{equation}
\label{eq:c}
c=\sqrt{\frac 53\bar\theta\left(1+\frac{8\bar\rho}{\rho_{sp}}
  +\frac{32\bar\rho^2}{5\rho_{sp}^2}\right)}.
\end{equation}
Setting $\bar\rho/\rho_{sp}=0$, we obtain the conventional speed of
sound in monatomic gases \cite{Gols}.

\begin{figure}%
  \begin{center}%
    \includegraphics[width=0.5\textwidth]{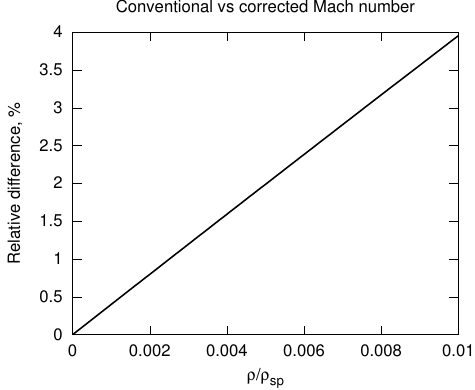}%
  \end{center}%
  \caption{The relative difference between the conventional and
    corrected Mach numbers as a function of $\bar\rho/\rho_{sp}$.}%
  \label{fig:Mach}%
\end{figure}

One can express the Enskog correction in the speed of sound by
introducing, respectively, the conventional $\Mach$ and
Enskog-corrected $\CMach$ Mach numbers for monatomic gases:
\begin{equation}
\label{eq:Mach}
\Mach=\frac{\bar u}{\displaystyle\sqrt{\frac 53\bar\theta}},\qquad
\CMach=\frac{\bar u}{\displaystyle\sqrt{\frac 53\bar\theta\left(1+
\frac{8\bar\rho}{\rho_{sp}}+\frac{32\bar\rho^2}{5\rho_{sp}^2}\right)}}.
\end{equation}
As we can see, for a given velocity $\bar u$, the corrected Mach
number $\CMach$ is somewhat smaller than the conventional one, because
the corresponding corrected speed of sound $c$ is somewhat larger than
the conventional one. We display the relative difference between the
conventional and corrected Mach numbers as a function of
$\bar\rho/\rho_{sp}$ in Figure~\ref{fig:Mach}.
Observe that the relative difference between the conventional and
corrected Mach numbers reaches 4\% for the density ratio
$\rho/\rho_{sp}=0.01$.

\section{The steady one-dimensional case: reduction to the system
of two ODEs}
\label{sec:steady}

In what follows, we examine nontrivial steady one-dimensional
solutions of the Enskog--Navier--Stokes equations. The imposed
conditions are quite stringent; in particular, such a solution implies
that the following conditions hold:
\begin{enumerate}
\item The solution is, in effect, a one-dimensional nonlinear
  nondispersive wave, and
\item The chosen reference frame moves at an appropriate speed, so
  that this wave is motionless in such a reference frame.
\end{enumerate}
In reality, shockwaves in gases do not behave exactly in this way, as
there is always some dispersion and/or dissipation present in
practical situations. However, the shockwaves behave substantially
similarly to such solutions, and thus the latter are still physically
meaningful. Also, as we will see below, such stringent requirements on
solutions allow to reduce the system of three partial differential
equations above into a system of two ordinary differential equations,
which is relatively simple to examine.

Assuming uniformity in the $t$-direction, we simplify
\eqref{eq:enskog_navier_stokes_1D} above into
\begin{subequations}
\begin{equation}
\deriv{}x(\rho u)=0,
\end{equation}
\begin{equation}
\deriv{}x\left(\rho u^2+p\right)=\frac 43\deriv{}x \left(\mu
\left(1+\frac 32a_1 -\frac 12a_2\right)\deriv ux \right),
\end{equation}
\begin{equation}
\deriv{}x\big((\rho\epsilon+p)u\big)=\frac{15}4\deriv{}x\left(\mu(1+a_3)
\deriv\theta x\right)+\frac 43\deriv{}x\left(\mu\left(1+ \frac 32a_1-
\frac 12a_2\right)u\deriv ux\right).
\end{equation}
\end{subequations}
Integrating on both sides, we arrive at
\begin{subequations}
\begin{equation}
\rho u=\rho_0u_0,
\end{equation}
\begin{equation}
\frac 43\mu\left(1+\frac 32a_1-\frac 12a_2\right)\deriv ux=\left(\rho
u^2+p\right)-\left(\rho_0 u_0^2+p_0\right),
\end{equation}
\begin{equation}
\frac{15}4\mu(1+a_3)\deriv\theta x+\frac 43\mu\left(1+\frac 32a_1-
\frac 12a_2\right)u\deriv ux=(\rho\epsilon+p)u-(\rho_0\epsilon_0+
p_0)u_0,
\end{equation}
\end{subequations}
where $\rho_0$, $u_0$, $\epsilon_0$ and $p_0$ denote the initial
values of the density, velocity, kinetic energy and pressure.  In
particular, if the terminal, post-shock values are defined via
$\rho_1$, $u_1$, $\epsilon_1$ and $p_1$, then it is easy to see that
the corresponding Rankine--Hugoniot shock conditions are given via
\begin{equation}
\rho_0 u_0=\rho_1 u_1,\qquad\rho_0 u_0^2+p_0=\rho_1 u_1^2+p_1,\qquad
\epsilon_0+p_0/\rho_0=\epsilon_1+p_1/\rho_1.
\end{equation}
Multiplying the second equation by $u$ and subtracting from the third,
we arrive at
\begin{subequations}
\begin{equation}
\rho u=\rho_0u_0,
\end{equation}
\begin{equation}
\frac 43\mu\left(1+\frac 32a_1-\frac 12a_2\right)\deriv ux=\left( \rho
u^2+p\right)-\left(\rho_0 u_0^2+p_0\right),
\end{equation}
\begin{equation}
\frac{15}4\mu(1+a_3)\deriv\theta x=\rho u(\epsilon-u^2)-\rho_0 u_0
(\epsilon_0-u_0 u)+p_0(u-u_0).
\end{equation}
\end{subequations}
In what follows, we write the equations above entirely via the $\rho,
u,\theta$ variables, since they are defined in a straightforward
fashion by the corresponding velocity moments in \eqref{eq:rho_u_e}
and \eqref{eq:theta_p}. The kinetic energy $\epsilon$ and pressure $p$
are expressed in terms of $\rho, u,\theta$ via \eqref{eq:theta_p}.
Expressing
\begin{equation}
u=\frac{\rho_0u_0}\rho,\qquad\deriv ux=-\frac{\rho_0u_0}{\rho^2}\deriv
\rho x,
\end{equation}
we exclude the velocity $u$ from the set of unknowns of the system,
and arrive at
\begin{subequations}
\label{eq:ODE}
\begin{equation}
\frac 43\mu\left(1+\frac 32a_1-\frac 12a_2\right)\deriv\rho x=\rho^2
u_0\left(1-\frac{\rho_0}\rho-\frac{\theta_0}{u_0^2}\left(\frac{\rho
  \theta}{\rho_0\theta_0}-1+\frac{4\rho_0}{\rho_{sp}}\left(\frac{\rho^2
  \theta}{\rho_0^2\theta_0}-1\right)\right)\right),
\end{equation}
\begin{equation}
\frac{15}2\mu(1+a_3)\deriv\theta x=\rho_0u_0\theta_0\left(3\left(
\frac\theta{\theta_0}-1\right)-\frac{u_0^2}{\theta_0}\left(1-\frac{
  \rho_0}\rho\right)^2-2\left(1+\frac{4 \rho_0}{\rho_{sp}}\right)
\left(1-\frac{\rho_0}\rho\right)\right).
\end{equation}
\end{subequations}
This is a $2\times 2$ system of ordinary differential equations, with
$\rho(x)$ and $\theta(x)$ being unknown functions, and $\rho_0$, $u_0$
and $\theta_0$ being constant parameters.

In what follows, we look for solutions of \eqref{eq:ODE} which
approach one fixed point at $x\to-\infty$, and another, distinct,
fixed point at $x\to+\infty$, with a transition in between. This
approach corresponds well to what is normally observed in shockwaves
traveling through a gas -- typically, there is a spatially uniform
state before the shock transition, and then the gas relaxes to another
spatially uniform state after the shock transition. Note that one of
the fixed points of the system in \eqref{eq:ODE} is already given via
$\rho(x)=\rho_0$, $\theta(x)=\theta_0$. Thus, we first need to find
other fixed points of \eqref{eq:ODE}, and then examine the orbits
which connect them.

\section{The fixed points}
\label{sec:fixed}

Above, we arrived at the system \eqref{eq:ODE} of two ODEs for $\rho$
and $\theta$ as functions of the spatial coordinate. First, we examine
the remaining fixed points of the system, in addition to the already
known fixed point $\rho(x)=\rho_0$, $\theta(x)=\theta_0$.

To compute the fixed points, we set the derivatives of $\rho$ and
$\theta$ above in \eqref{eq:ODE} to zero, and observe that both
resulting equations are linear in $\theta$. This allows us to express
$\theta$ via $\rho$ with help of the second equation in \eqref{eq:ODE}:
\begin{equation}
\theta=\theta_0+\frac 13\left(1-\frac{\rho_0}\rho\right)\left[u_0^2
  \left(1-\frac{\rho_0}\rho\right)+2\theta_0\left(1+\frac{4\rho_0
  }{\rho_{sp}}\right)\right].
\end{equation}
Next, we substitute the expression above into the first equation in
\eqref{eq:ODE}, which results in following algebraic equation for
$\rho$ alone:
\begin{multline}
-\rho_0u_0^2(\rho-\rho_0)+\rho\theta_0(\rho-\rho_0)+\frac 4{
  \rho_{sp}}\rho\theta_0(\rho-\rho_0)(\rho+\rho_0)+\\+\frac 13\left(
1+\frac{4\rho}{\rho_{sp}}\right)\rho\left(\rho-\rho_0\right)\left(
u_0^2\left(1-\frac{\rho_0}\rho\right)+2\theta_0\left(1+\frac{4\rho_0}{
  \rho_{sp}}\right)\right)=0.
\end{multline}
Clearly, one of the solutions is $\rho=\rho_0$, which was already
mentioned above. To look for a different root, we divide by
$(\rho-\rho_0)$, and arrive at the quadratic equation
\begin{equation}
\label{eq:rho2}
\frac 4{\rho_{sp}}\left(u_0^2+5\theta_0\left(1+\frac{8\rho_0}{5
  \rho_{sp}}\right)\right)\rho^2+\left(5\theta_0\left(1+\frac {4
  \rho_0}{\rho_{sp}}\right)+u_0^2\left(1-\frac {4\rho_0}{\rho_{sp}}
\right)\right)\rho-4\rho_0u_0^2=0.
\end{equation}
This equation has two real roots -- one positive, and one negative.
We, however, are interested only in the positive root, since $\rho$ is
the density and cannot be negative. The positive root of
\eqref{eq:rho2} is given via
\begin{equation}
\label{eq:rho1}
\rho_1=\frac{\rho_{sp}}8\frac{3+\Mach_0^2\left(1-\frac{4\rho_0}{
    \rho_{sp}}\right)+\frac{12\rho_0}{\rho_{sp}}}{3+\Mach_0^2+\frac{
    24\rho_0}{5\rho_{sp}}}\left(\sqrt{1+\frac{\rho_0}{\rho_{sp}}
  \frac{64\Mach_0^2\left(3+\Mach_0^2+\frac{24\rho_0}{5\rho_{sp}}
    \right)}{\left( 3 + \Mach_0^2 \left(1-\frac{ 4\rho_0}{\rho_{sp}}
    \right)+\frac{12\rho_0}{\rho_{sp}} \right)^2}}-1\right).
\end{equation}
Knowing $\rho_1$, we can express $u_1$ and $\theta_1$ via
\begin{equation}
\label{eq:u1theta1}
u_1=\frac{\rho_0u_0}{\rho_1},\qquad\theta_1=\theta_0\left(1+\frac
59\left(1
-\frac{\rho_0}{\rho_1}\right)\left(\frac 65+\Mach_0^2\left(1-\frac{\rho_0}{\rho_1}
\right)+\frac{24\rho_0}{5\rho_{sp}}\right)\right).
\end{equation}
Combining \eqref{eq:rho1} and \eqref{eq:u1theta1} together, we can write
\begin{equation}
(\rho_1,u_1,\theta_1)=\cF_{\rho_{sp}}(\rho_0,u_0,\theta_0).
\end{equation}
Clearly, $\cF_{\rho_{sp}}\circ\cF_{\rho_{sp}}=$ identity. In what
follows, we will tacitly assume that $(\rho_0,u_0,\theta_0)$
corresponds to the state before the shock (that is, the one with
higher velocity and lower density and temperature), while
$(\rho_1,u_1,\theta_1)$ corresponds to the state after the shock (the
one with lower velocity and higher density and temperature), unless
otherwise specified.

To find the conditions under which the fixed points coalesce (that is,
the state before the shock is identical to the state after the shock),
we equate $\rho_1=\rho_0$ in \eqref{eq:rho1}. This yields
\begin{equation}
u_0^2=\frac 53\theta_0\left(1+\frac{8\rho_0}{\rho_{sp}}+\frac{32
  \rho_0^2}{5\rho_{sp}^2}\right),
\end{equation}
that is, the Enskog-corrected Mach number $\CMach_0$ in
\eqref{eq:Mach}, corresponding to $\rho_0$, $u_0$ and $\theta_0$, must
be equal to 1, which is to be expected.

\subsection{The fixed points for the conventional Navier--Stokes
equations}

To obtain the fixed points for the conventional Navier--Stokes
equations, we set $\rho_0/\rho_{sp}=0$ above in \eqref{eq:rho1} and
\eqref{eq:u1theta1}. This yields the following, much more simple,
relations
\begin{equation}
\rho_{1,NS}=\rho_0\frac{4\Mach_0^2}{\Mach_0^2+3},\qquad
u_{1,NS}=u_0\frac{\Mach_0^2+3}{4\Mach_0^2},\qquad
\theta_{1,NS}=\theta_0\frac{(\Mach_0^2+3)(5\Mach_0^2-1)}{16\Mach_0^2}.
\end{equation}
Here we can see that the conventional Mach number cannot be less than
$1/\sqrt 5$ (otherwise, the temperature would become negative). We can
also conveniently express the post-shock Mach number in terms of the
pre-shock Mach number via
\begin{equation}
\Mach_1^2=\frac 35\frac{u_{1,NS}^2}{\theta_{1,NS}}=\frac{\Mach_0^2+3}{5
  \Mach_0^2-1}.
\end{equation}

\subsection{Qualitative behavior for a small density ratio}

For a small density ratio $\rho_0/\rho_{sp}$, we expand
\eqref{eq:rho1} and \eqref{eq:u1theta1} in leading order to obtain
\begin{subequations}
\label{eq:linear_approx}
\begin{equation}
\rho_1\approx\rho_0\frac{4\Mach_0^2}{\Mach_0^2+3}\left(1-\frac{12\rho_0}{
  \rho_{sp}}\frac{\Mach_0^2+1}{\Mach_0^2+3}\right),
\end{equation}
\begin{equation}
u_1\approx u_0\frac{\Mach_0^2+3}{4\Mach_0^2}\left(1+\frac{12\rho_0}{
  \rho_{sp}}\frac{\Mach_0^2+1}{\Mach_0^2+3}\right).
\end{equation}
\begin{equation}
\theta_1\approx\theta_0\frac{(\Mach_0^2+3)(5\Mach_0^2-1)}{16\Mach_0^2}
\left(1-\frac{8\rho_0}{ \rho_{sp}}\frac{5\Mach_0^4+3}{(\Mach_0^2+3)(
  5\Mach_0^2-1)}\right).
\end{equation}
\end{subequations}
Observe that the correction factor always reduces the difference
between the pre-shock and post-shock states. Assuming that $\rho_0$,
$u_0$ and $\theta_0$ are the values corresponding to the region before
the shock (i.e. $\Mach_0>1$), we conclude that the relative effect of
the correction factor is weakest for $\Mach_0\approx 1$, and increases
as $\Mach_0\to\infty$.
\begin{figure}%
\includegraphics[width=0.5\textwidth]{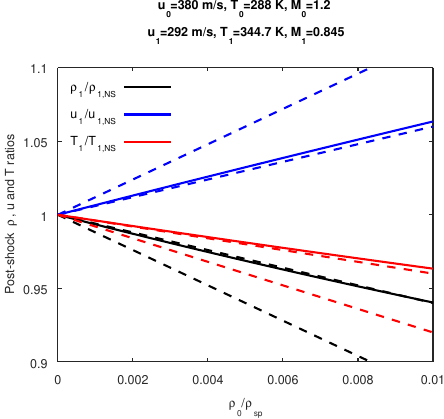}%
\includegraphics[width=0.5\textwidth]{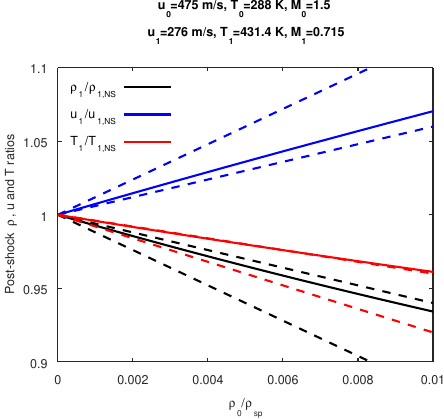}\\%
\includegraphics[width=0.5\textwidth]{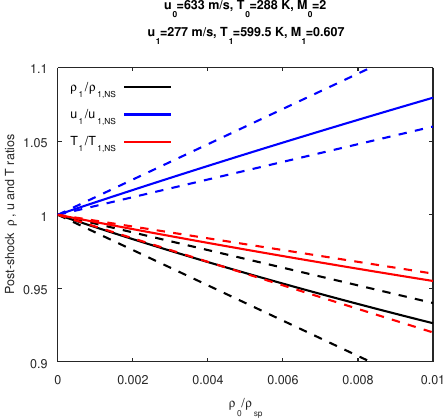}%
\includegraphics[width=0.5\textwidth]{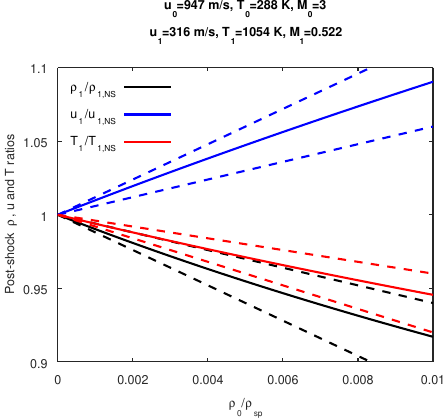}%
\caption{Post-shock ratios $\rho/\rho_{NS}$, $u/u_{NS}$, and
  $T/T_{NS}$ for different incident Mach numbers. The pre-shock
  temperature is set to 288 K, and corresponds to standard
  conditions.}
\label{fig:shocks_fixed}
\end{figure}

To illustrate the effect of the Enskog correction on the fixed points,
in Figure \ref{fig:shocks_fixed} we show the post-shock ratios
$\rho_1/\rho_{1,NS}$, $u_1/u_{1,NS}$, $T_1/T_{1,NS}$ (where we display
the conventional temperature $T$ instead of $\theta$ to provide a
better understanding of the thermodynamic regime, see
\eqref{eq:temperature} for the relation between $\theta$ and $T$) for
the given pre-shock values of $\rho_0/\rho_{sp}$, $u_0$ and $T_0$ of
an argon-like hard sphere gas, with the physical parameters of the
spheres specified as follows:
\begin{itemize}
\item The sphere diameter $\sigma=3.6\cdot 10^{-10}$ meters;
\item The sphere density $\rho_{sp}=2716$ kg/m$^3$.
\end{itemize}
Such a sphere possesses roughly the same mass as one atom of argon,
and has viscosity $\mu\approx 2.2\cdot 10^{-5}$ kg/(m s$^2$) at
$T=288$ K, which is also similar to argon. Together with the exactly
computed (via \eqref{eq:rho1} and \eqref{eq:u1theta1}) ratios, which
are given as solid lines, we also plot the linearized cones given via
\eqref{eq:linear_approx} for minimal ($1$) and maximal ($\infty$)
pre-shock Mach number values as dashed lines of the same color. The
pre-shock temperature is set at $T_0=288$ K, while the pre-shock
velocity $u_0$ is chosen to represent the four incident Mach number
regimes, $\Mach_0=1.2$, $1.5$, $2$ and $3$. The range of the pre-shock
density values $\rho_0$ is chosen so that the ratio $\rho_0/\rho_{sp}$
varies between zero and 0.01. Observe that, for the presented range of
$\rho_0/\rho_{sp}$, the exactly computed ratios $\rho_1/\rho_{1,NS}$,
$u_1/u_{1,NS}$, $T_1/T_{1,NS}$ behave almost linearly, and generally
fall within the linearized cones. We can observe that, for
$\rho_0/\rho_{sp}=0.01$, the effect of the Enskog correction varies
between roughly 5\% (for transonic Mach numbers) to about 10\% (for
supersonic Mach numbers). For the normal atmospheric pressure and
temperature, the equivalent argon density is about $1.7$ kg/m$^3$,
which corresponds to roughly 1\% of the effect of the Enskog
correction.

\section{The shock transitions}
\label{sec:transitions}

With the information about the relevant fixed points of
\eqref{eq:ODE}, we can now study the transition between them. Clearly,
if we accept that the state of the gas before and after the shock is
modeled via $(\rho_0,u_0,\theta_0)$ and $(\rho_1,u_1,\theta_1)$,
respectively, then the corresponding shock transition must be given
via the heteroclinic orbit\cite{GucHol} of \eqref{eq:ODE} which
connects these two fixed points.

Due to complexity of \eqref{eq:ODE}, we have to compute this
heteroclinic orbit numerically. This is possible to do, with good
accuracy, in the following scenario:
\begin{itemize}
\item One fixed point is a stable node;
\item Another fixed point is a saddle node;
\item The heteroclinic orbit is one of the two branches of the
  unstable manifold of the saddle node.
\end{itemize}
In such a case, the heteroclinic orbit will be a local attracting set
for nearby orbits, while the stable-node fixed point will be a local
attractor. Therefore, if we choose the initial condition sufficiently
close to the saddle-node fixed point, then its orbit will shrink
transversally towards the heteroclinic orbit due to the fact that the
stable manifold of the saddle-node fixed point is transversal to the
heteroclinic orbit, while at the same time stretching along the
unstable manifold and eventually falling onto the local attractor. In
other words, the accuracy of the solution will be ensured by the fact
that the remaining invariant manifold of the saddle node is stable.
The same idea for the computation of such a shock transition was
suggested in \cite{GilPao} for the conventional Navier--Stokes
equations. In \cite{GilPao}, the heteroclinic orbit is referred to as
the ``shock curve''.

We start by introducing the nondimensional notations
\begin{equation}
r = \frac\rho{\rho_0},\qquad q=\frac\theta{\theta_0},\qquad \mu=\eta
q^{1/2},\qquad\eta=\frac{5\sqrt\pi \rho_{sp}\sigma\sqrt
  \theta_0}{96},\qquad x=\frac\eta{\rho_0u_0}z,
\end{equation}
so that $r_0=1$, $q_0=1$. This allows us to write \eqref{eq:ODE} in
the nondimensional form
\begin{subequations}
\label{eq:ODE2}
\begin{equation}
\deriv rz=F(r,q),\qquad\deriv qz=G(r,q),
\end{equation}
\begin{equation}
F=\frac 34\left(1+\frac 32a_1-\frac 12a_2\right)^{-1}q^{-1/2}\left(r
(r-1)-\frac{\theta_0}{u_0^2}r^2\left((rq-1)+\frac{4\rho_0}{\rho_{sp}}
(r^2q-1)\right)\right),
\end{equation}
\begin{equation}
G=\frac 2{15}(1+a_3)^{-1}q^{-1/2}\left(3(q-1)-\frac{u_0^2}{\theta_0}
\frac 1{r^2}\left(r-1\right)^2-2\left(1+\frac{4\rho_0}{ \rho_{sp}}
\right)\frac 1r\left(r-1\right)\right).
\end{equation}
\end{subequations}
Here we observe that
\begin{equation}
  3a_1-a_2>0,\qquad a_3>0,
\end{equation}
which means that the vector field above is smooth everywhere in the
first quadrant of the $(r,q)$-plane.

\begin{figure}%
\includegraphics[width=\textwidth]{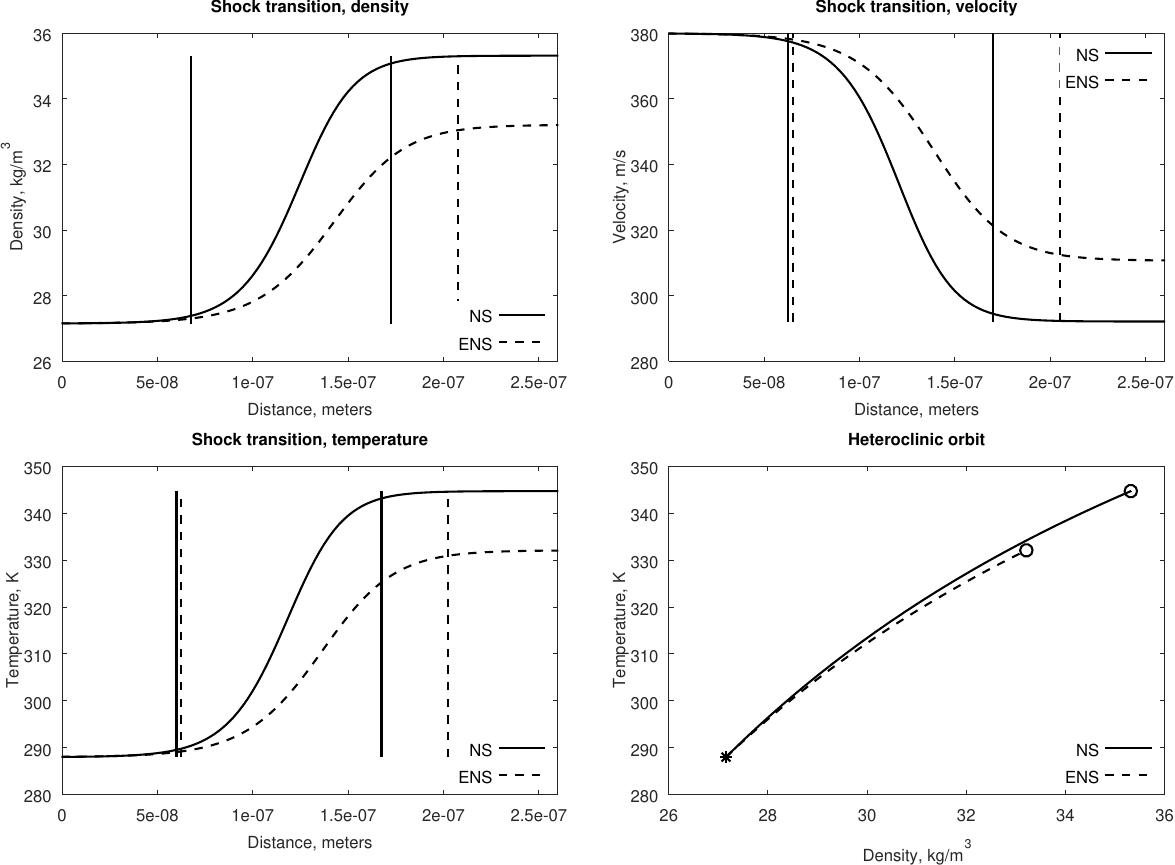}%
\caption{The shock transition for $\rho_0=27.16$ kg/m$^3$, $u_0=380$
  m/s, $T_0=288$ K, $\Mach_0=1.2$. Solid line -- conventional
  Navier--Stokes transition, dashed line -- Enskog-corrected
  Navier--Stokes transition. The boundaries of the shock transitions
  are marked by the vertical lines of corresponding styles.}%
\label{fig:shock_transition_u380}%
\end{figure}%
\begin{figure}%
\includegraphics[width=0.7\textwidth]{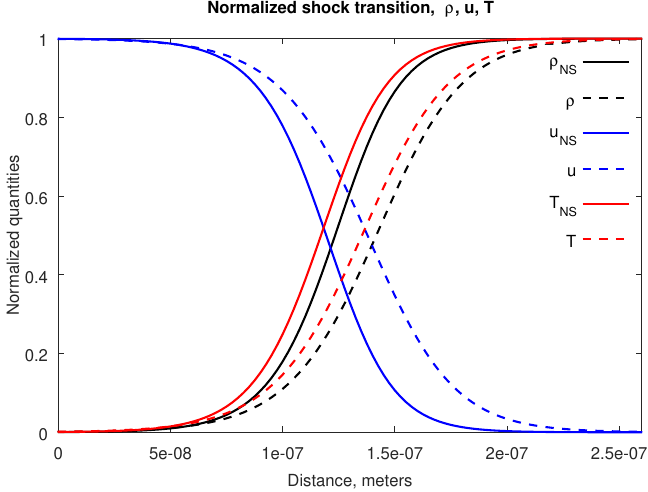}%
\caption{The normalized shock transition for $\rho_0=27.16$ kg/m$^3$, $u_0=380$
  m/s, $T_0=288$ K, $\Mach_0=1.2$. Solid line -- conventional
  Navier--Stokes transition, dashed line -- Enskog-corrected
  Navier--Stokes transition.}%
\label{fig:shock_transition_normalized_u380}%
\end{figure}%
Now, we need to determine the stability of the fixed point $(1,1)$.
This will be accomplished via the Grobman--Hartman theorem
\cite{Gro,Har}; if the fixed point is indeed hyperbolic, then its
stability properties are the same as those of the linearized
system. For that, we compute the partial derivatives
\begin{subequations}
\begin{equation}
\parderiv Fr(1,1)=\frac 34\left(1+\frac 32a_1-\frac 12a_2\right)^{-1}
\left(1-\frac{\theta_0}{u_0^2}\left(1+\frac{8\rho_0}{\rho_{sp}}
\right)\right),
\end{equation}
\begin{equation}
\parderiv Fq(1,1)=-\frac 34\left(1+\frac 32a_1-\frac
12a_2\right)^{-1}\frac{\theta_0}{u_0^2}\left(1+\frac{4\rho_0}{\rho_{sp}}
\right),
\end{equation}
\begin{equation}
\parderiv Gr(1,1)=-\frac 4{15}(1+a_3)^{-1}\left(1+\frac{4\rho_0
}{\rho_{sp}}\right),\qquad\parderiv Gq(1,1)=\frac 25(1+a_3)^{-1},
\end{equation}
\end{subequations}
where, of course, $a_1$, $a_2$ and $a_3$ are evaluated for
$\rho=\rho_0$.  The characteristic equation is given via
\begin{equation}
\lambda^2-\left(\parderiv Fr(1,1)+\parderiv Gq(1,1)\right)
\lambda+\parderiv Fr(1,1)\parderiv Gq(1,1)-\parderiv Fq(1,1)\parderiv
Gr(1,1)=0,
\end{equation}
which has roots of the same sign if
\begin{equation}
\parderiv Fr(1,1)\parderiv Gq(1,1)-\parderiv Fq(1,1)\parderiv
Gr(1,1)>0,
\end{equation}
and roots of the opposite sign otherwise. Using the expressions above,
we find that, for the roots of the same sign, we need to have
\begin{equation}
\frac 53\left(1+\frac{8\rho_0}{\rho_{sp}} +\frac{32\rho_0^2}{
  5\rho_{sp}^2}\right)\frac{\theta_0}{u_0^2}<1,
\end{equation}
or, alternatively,
\begin{equation}
\CMach_0>1.
\end{equation}
\begin{figure}%
\includegraphics[width=\textwidth]{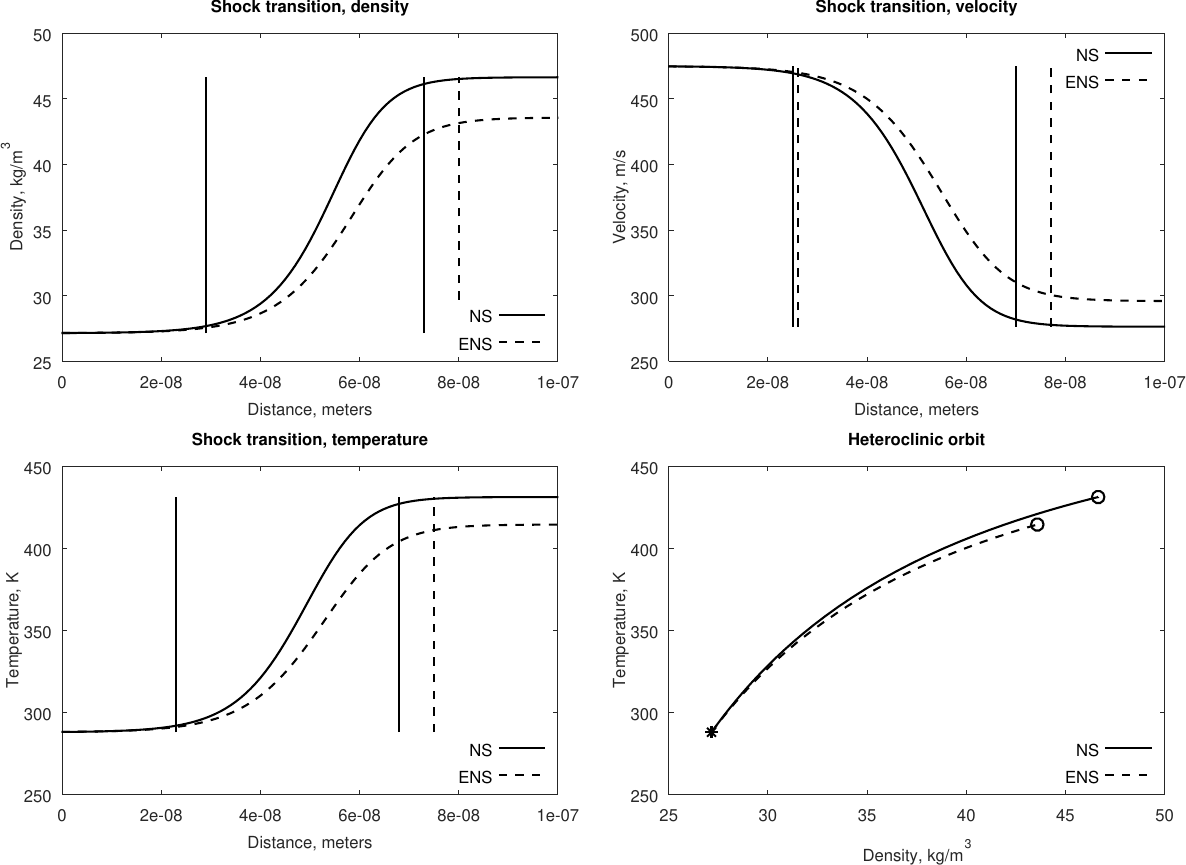}%
\caption{The shock transition for $\rho_0=27.16$ kg/m$^3$, $u_0=475$
  m/s, $T_0=288$ K, $\Mach_0=1.5$. Solid line -- conventional
  Navier--Stokes transition, dashed line -- Enskog-corrected
  Navier--Stokes transition. The boundaries of the shock transitions
  are marked by the vertical lines of corresponding styles.}%
\label{fig:shock_transition_u475}%
\end{figure}%
\begin{figure}%
\includegraphics[width=0.7\textwidth]{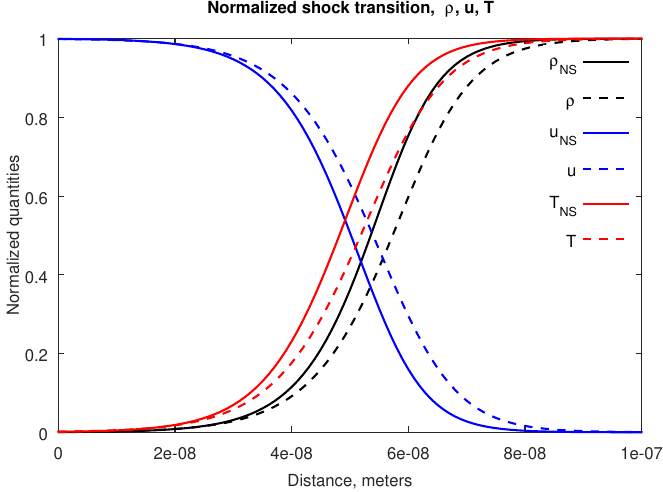}%
\caption{The normalized shock transition for $\rho_0=27.16$ kg/m$^3$, $u_0=475$
  m/s, $T_0=288$ K, $\Mach_0=1.5$. Solid line -- conventional
  Navier--Stokes transition, dashed line -- Enskog-corrected
  Navier--Stokes transition.}%
\label{fig:shock_transition_normalized_u475}%
\end{figure}%
Observing that for $\CMach_0>1$ both partial derivatives $\partial
F/\partial r$ and $\partial G/\partial q$ are positive, we determine
that both roots are positive, which means that the pre-shock fixed
point is an unstable node, and, therefore, the post-shock fixed point
is a saddle.  The same analysis, but for the conventional
Navier--Stokes equations, is presented in Chapter 12 of
\cite{MulRug2}. Thus, in order to capture the heteroclinic orbit (and
thus the shock transition), we integrate \eqref{eq:ODE2}
``backwards'', that is, from the post-shock state
$(\rho_1,u_1,\theta_1)$ towards the pre-shock state
$(\rho_0,u_0,\theta_0)$, which effectively reverts the sign of $z$ in
\eqref{eq:ODE2}, rendering the pre-shock state attracting.

\begin{figure}%
\includegraphics[width=\textwidth]{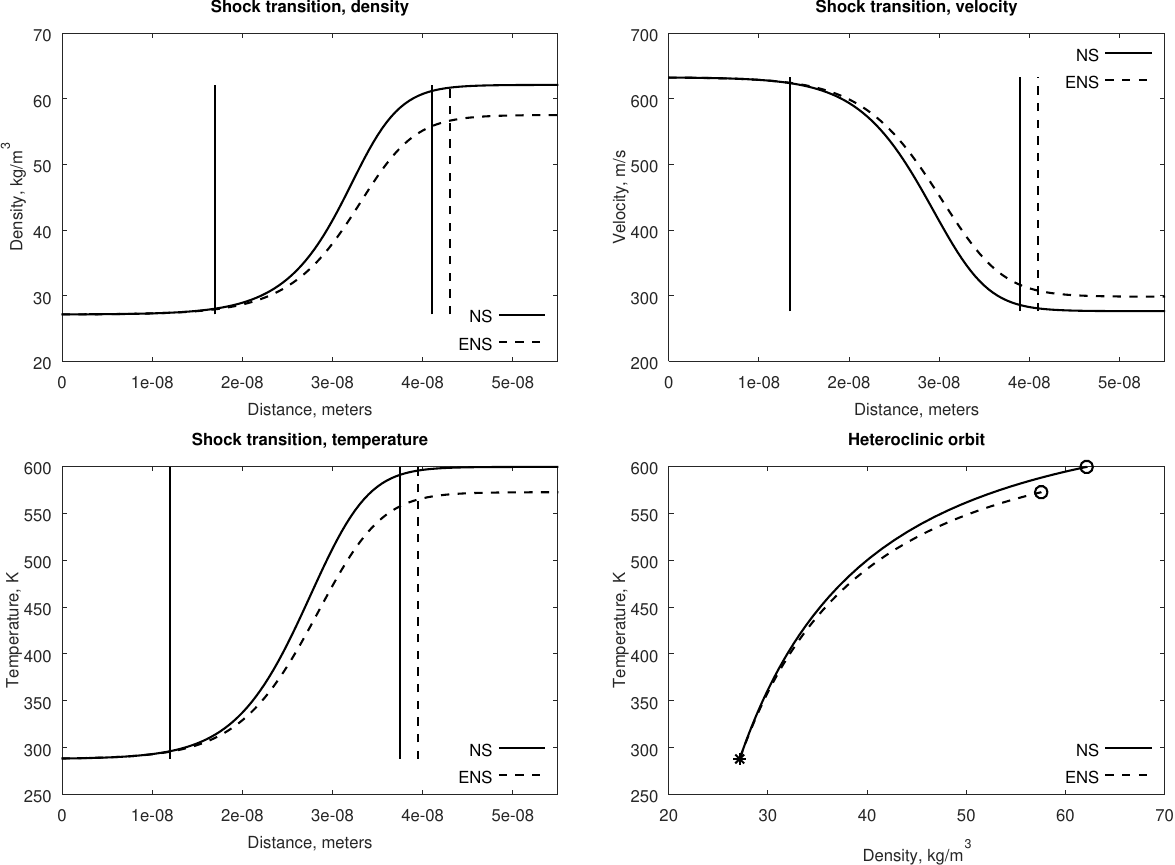}%
\caption{The shock transition for $\rho_0=27.16$ kg/m$^3$, $u_0=633$
  m/s, $T_0=288$ K, $\Mach_0=2$. Solid line -- conventional
  Navier--Stokes transition, dashed line -- Enskog-corrected
  Navier--Stokes transition. The boundaries of the shock transitions
  are marked by the vertical lines of corresponding styles.}
\label{fig:shock_transition_u633}
\end{figure}
\begin{figure}%
\includegraphics[width=0.7\textwidth]{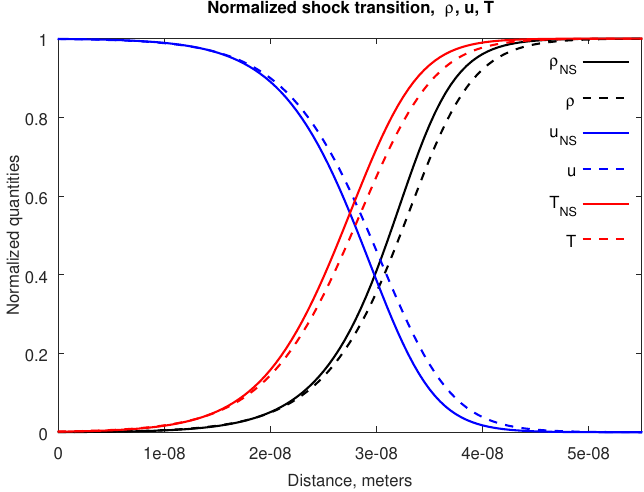}%
\caption{The normalized shock transition for $\rho_0=27.16$ kg/m$^3$, $u_0=633$
  m/s, $T_0=288$ K, $\Mach_0=2$. Solid line -- conventional
  Navier--Stokes transition, dashed line -- Enskog-corrected
  Navier--Stokes transition.}
\label{fig:shock_transition_normalized_u633}
\end{figure}
\begin{figure}%
\includegraphics[width=\textwidth]{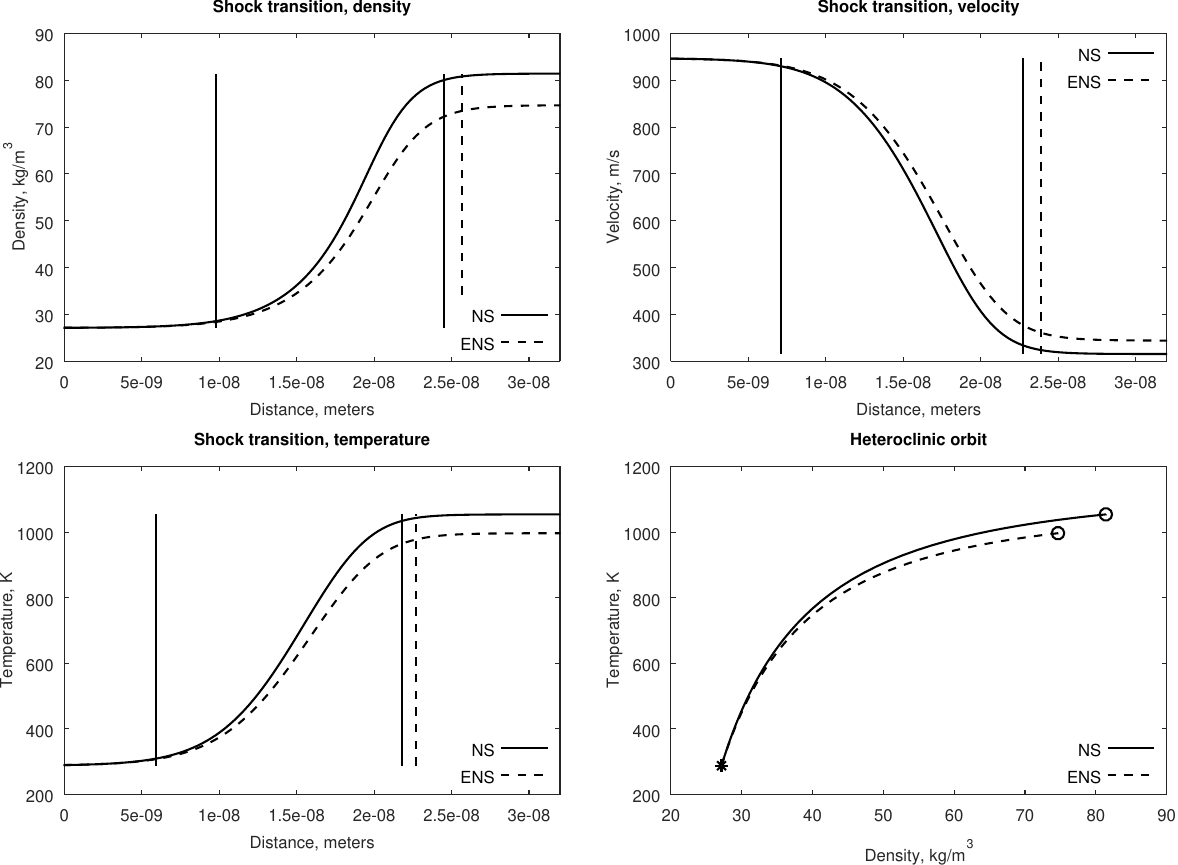}%
\caption{The shock transition for $\rho_0=27.16$ kg/m$^3$, $u_0=947$
  m/s, $T_0=288$ K, $\Mach_0=3$. Solid line -- conventional
  Navier--Stokes transition, dashed line -- Enskog-corrected
  Navier--Stokes transition. The boundaries of the shock transitions
  are marked by the vertical lines of corresponding styles.}
\label{fig:shock_transition_u947}
\end{figure}
\begin{figure}%
\includegraphics[width=0.7\textwidth]{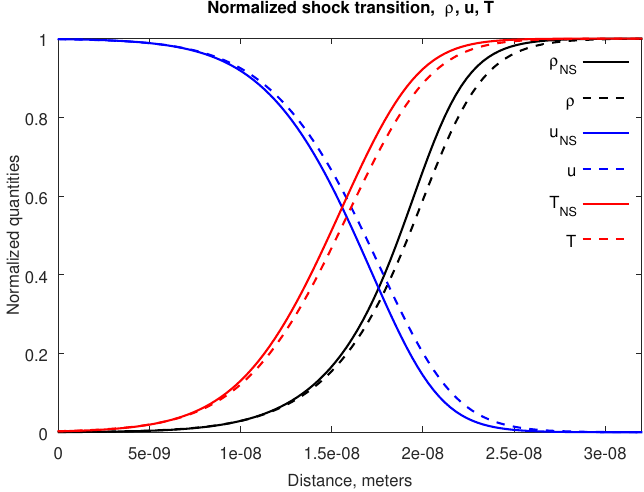}%
\caption{The normalized shock transition for $\rho_0=27.16$ kg/m$^3$, $u_0=947$
  m/s, $T_0=288$ K, $\Mach_0=3$. Solid line -- conventional
  Navier--Stokes transition, dashed line -- Enskog-corrected
  Navier--Stokes transition.}
\label{fig:shock_transition_normalized_u947}
\end{figure}

In Figures
\ref{fig:shock_transition_u380}--\ref{fig:shock_transition_normalized_u947}
we compute the shock transitions, as described above, for $\rho$, $u$,
and the conventional temperature $T$ (see \eqref{eq:temperature} for
the relation between $\theta$ and $T$) for the same argon-like hard
sphere gas as above. To numerically capture the heteroclinic orbit in
\eqref{eq:ODE2}, we perturb the post-shock fixed point by $10^{-6}$\%
in the direction of the pre-shock fixed point, and then use the
\textsl{lsode} routine \cite{Hin} of the Octave software \cite{Octave}
to carry out the numerical integration.  We compute the spatial
``thickness'' of the shock transition as the length on which 95\% of
the transition occurs, situated symmetrically between the fixed points
-- that is, the transition starts when 2.5\% of the distance between
the fixed points is covered, and ends when 2.5\% of the distance
remains. The start and end of each transition are marked on plots by
vertical lines.

In all cases, we set $T_0=288$ K and $\rho_0=27.16$ kg/m$^3$ (so that
$\rho_0/\rho_{sp}=0.01$), and vary $u_0$ to achieve different incident
Mach numbers. This set-up corresponds to a moderately dense gas --
given that, at the sea-level pressure, the density of argon is about
$1.7$ kg/m$^3$, the equivalent density in our set-up is roughly 16
Earth atmospheres. This is a physically realistic scenario -- for
example, the equivalent density for the same gas in the Cytherean
atmosphere corresponds to the altitude of about 20-30 kilometers above
surface, while much greater densities can be achieved in the
respective atmospheres of the four gas giants of the Solar
system. Moreover, the density of the gaseous part of a propellant in
the combustion chamber of a rocket engine can also be of the same
order of magnitude.

In Figures \ref{fig:shock_transition_u380} and
\ref{fig:shock_transition_normalized_u380} we compute the shock
transition for $u_0=380$ m/s, or $\Mach_0=1.2$. Figure
\ref{fig:shock_transition_u380} shows the transitions in density,
velocity and temperature separately in their respective physical
units, while in Figure \ref{fig:shock_transition_normalized_u380} we
normalize all transitions so that they vary between 0 and 1, and
overlay them on the same plot. In Figure
\ref{fig:shock_transition_u380}, we also display the heteroclinic
orbit between the pre-shock and post-shock fixed points in the
$(\rho,T)$-phase plane. Observe that, apart from the significant
difference between the post-shock states for the Navier--Stokes and
Enskog-corrected Navier--Stokes dynamics (which was already predicted
in the previous section), there is another interesting effect of the
Enskog correction -- the transition between the pre-shock and
post-shock states for the Enskog-corrected dynamics is longer and
smoother than that for the conventional Navier--Stokes equations, so
that the ``thickness'' of the shock is greater. We note that a similar
behavior was observed in \cite{GreRee}, which was attributed to an
additional proposed mass-diffusion effect in gases
\cite{Brenner,Brenner2,Brenner3,DadRee,Durst,Durst2,GreRee}. We do
not, however, insist that what we observe here and what was observed
in \cite{GreRee} is necessarily the same effect, but merely point out
the qualitative similarity. Also, observe that the heteroclinic orbit
in the $(\rho,T)$-phase plane, shown in Figure
\ref{fig:shock_transition_u380}, is close to a straight line (albeit
slightly concave), which means that the density and temperature shock
transitions occur almost synchronously, with temperature slightly
preceding the density (also see Figure
\ref{fig:shock_transition_normalized_u380} with overlaid normalized
plots).

In Figures \ref{fig:shock_transition_u475} and
\ref{fig:shock_transition_normalized_u475} we compute the shock
transition for $u_0=475$ m/s, or $\Mach_0=1.5$. Figure
\ref{fig:shock_transition_u475} shows the transitions in density,
velocity and temperature separately in their respective physical
units, while in Figure \ref{fig:shock_transition_normalized_u475} we
normalize all transitions so that they vary between 0 and 1, and
overlay them on the same plot. In Figure
\ref{fig:shock_transition_u475}, we also display the heteroclinic
orbit between the pre-shock and post-shock fixed points in the
$(\rho,T)$-phase plane. Here, observe that the relative difference of
the thickness of the shock transition between the conventional
Navier--Stokes equations and the Enskog-corrected dynamics diminishes
somewhat in comparison to the previous scenario with $\Mach_0=1.2$.
At the same time, the heteroclinic orbit in the $(\rho,T)$-phase
plane, shown in Figure \ref{fig:shock_transition_u475}, becomes more
concave, which means that the temperature shock transition precedes
the density shock transition a little more, relative to the overall
thickness of the shock, than in the previous scenario with
$\Mach_0=1.2$ (this can also be observed in Figure
\ref{fig:shock_transition_normalized_u475} with overlaid normalized
plots).

In Figures \ref{fig:shock_transition_u633} and
\ref{fig:shock_transition_normalized_u633} we compute the shock
transition for $u_0=633$ m/s, or $\Mach_0=2$. Unlike the two previous
scenarios with $\Mach_0=1.2$ and $\Mach_0=1.5$, which could both be
qualified as transonic or borderline transonic/supersonic, this is a
strongly supersonic scenario, where the speed of the shock wave is
roughly twice the speed of sound in the unperturbed gas in the
pre-shock zone. As before, we present two separate figures, with
Figure \ref{fig:shock_transition_u633} showing the transitions in
density, velocity and temperature separately in their respective
physical units, while Figure
\ref{fig:shock_transition_normalized_u633} displaying the normalized
transitions, overlaid on the same plot. In Figure
\ref{fig:shock_transition_u633}, we also display the heteroclinic
orbit between the pre-shock and post-shock fixed points in the
$(\rho,T)$-phase plane. Here, observe that the earlier observed shock
thickness trend continues -- the relative difference of the thickness
of the shock transition between the conventional Navier--Stokes
equations and the Enskog-corrected dynamics is smaller than that for
both previously examined Mach numbers, $\Mach_0=1.2$ and
$\Mach_0=1.5$. The heteroclinic orbit in the $(\rho,T)$-phase plane,
shown in Figure \ref{fig:shock_transition_u633}, becomes even more
concave in comparison with $\Mach_0=1.5$, which results in the
temperature shock transition preceding the density shock transition
even more, relative to the thickness of the shock itself, than in the
previous scenario with $\Mach_0=1.5$. The normalized plots in Figure
\ref{fig:shock_transition_normalized_u633} confirm the shift between
the density and temperature shock transition.

In Figures \ref{fig:shock_transition_u947} and
\ref{fig:shock_transition_normalized_u947} we compute the shock
transition for $u_0=947$ m/s, or $\Mach_0=3$. Like the previous
scenario with $\Mach_0=2$, this is also a strongly supersonic
scenario, where the speed of the shock wave is roughly thrice the
speed of sound in the unperturbed gas in the pre-shock zone. As
before, we present two separate figures, with Figure
\ref{fig:shock_transition_u947} showing the transitions in density,
velocity and temperature separately in their respective physical
units, while Figure \ref{fig:shock_transition_normalized_u947}
displaying the normalized transitions, overlaid on the same plot. In
Figure \ref{fig:shock_transition_u947}, we also display the
heteroclinic orbit between the pre-shock and post-shock fixed points
in the $(\rho,T)$-phase plane. Here, observe that the relative
difference in shock thicknesses between the conventional
Navier--Stokes equations and the Enskog-corrected dynamics is roughly
the same as for the previous scenario with $\Mach_0=2$ regime. The
heteroclinic orbit in the $(\rho,T)$-phase plane, shown in Figure
\ref{fig:shock_transition_u947}, is more concave than the one in the
$\Mach_0=2$ regime, and, as a result, the temperature shock transition
precedes the density shock transition even more, relative to the
thickness of the shock itself. The shift between the density and
temperature shock transitions is visible on the normalized plots in
Figure \ref{fig:shock_transition_normalized_u947}, for both the
conventional Navier--Stokes and Enskog-corrected dynamics.

We summarize the shock transition thicknesses and magnitudes in Tables
\ref{tab:thickness} and \ref{tab:magnitudes}, respectively. The tables
generally confirm the trends observed in the plots -- the relative
impact of the Enskog--corrected terms is greatest at the low Mach
numbers, where the difference in transition thicknesses can reach as
much as 30\%. Interestingly, there is not much of a relative
difference between $\Mach=2$ and $\Mach=3$ scenarios -- the relative
difference in transition thicknesses in both cases is around 8\%.

Here, we feel compelled to point out that the results in Table
\ref{tab:magnitudes} do not contradict the results in Figure
\ref{fig:shocks_fixed} (although it may seem so at a glance). Note
that the plots in Figure~\ref{fig:shocks_fixed} show the relative
effect of the Enskog correction on the post-shock state itself, while
the results in Table \ref{tab:magnitudes} show its effect on the {\em
  transition} between the pre-shock and post-shock state. So, while
the relative difference in post-shock states grows with the Mach
number (this trend is shown in Figure \ref{fig:shocks_fixed}), the
relative differences in shock transitions are more prominent at low
Mach numbers (as follows from Table \ref{tab:magnitudes}).
\begin{table}
\begin{tabular}{|c||c|c|}
\hline
& $\Mach_0=1.2$ & $\Mach_0=1.5$ \\
\hline
\begin{tabular}{c} \\ \hline NS \\ ENS \\ Diff. \end{tabular}
&
\begin{tabular}{c|c|c}
$\rho$ & $u$ & $T$ \\
\hline
$1.05 \cdot 10^{-7}$ & $1.075 \cdot 10^{-7}$ & $1.075 \cdot 10^{-7}$ \\
    $1.4 \cdot 10^{-7}$ & $1.4 \cdot 10^{-7}$ & $1.4 \cdot 10^{-7}$ \\
    $33$\% & $30$\% & $30$\%
\end{tabular} &
\begin{tabular}{c|c|c}
$\rho$ & $u$ & $T$ \\
\hline
$4.4 \cdot 10^{-8}$ & $4.5 \cdot 10^{-8}$ & $4.5 \cdot 10^{-8}$ \\
$5.1 \cdot 10^{-8}$ & $5.1 \cdot 10^{-8}$ & $5.2 \cdot 10^{-8}$ \\
$16$\% & $13$\% & $16$\%
\end{tabular}\\
\hline\hline
& $\Mach_0=2$ & $\Mach_0=3$ \\
\hline
\begin{tabular}{c} \\ \hline NS \\ ENS \\ Diff. \end{tabular}
&
\begin{tabular}{c|c|c}
$\rho$ & $u$ & $T$ \\
\hline
$2.4 \cdot 10^{-8}$ & $2.55 \cdot 10^{-8}$ & $2.55 \cdot 10^{-8}$ \\
$2.6 \cdot 10^{-8}$ & $2.75 \cdot 10^{-8}$ & $2.75 \cdot 10^{-8}$ \\
$8$\% & $8$\% & $8$\%
\end{tabular} &
\begin{tabular}{c|c|c}
$\rho$ & $u$ & $T$ \\
\hline
$1.47 \cdot 10^{-8}$ & $1.56 \cdot 10^{-8}$ & $1.59 \cdot 10^{-8}$ \\
$1.59 \cdot 10^{-8}$ & $1.68 \cdot 10^{-8}$ & $1.68 \cdot 10^{-8}$ \\
$8$\% & $8$\% & $6$\%
\end{tabular}\\
\hline
\end{tabular}
\caption{Thickness of shock transitions (meters) for all studied
  scenarios, as well as their relative differences in \%.}
\label{tab:thickness}
\end{table}

\begin{table}
\begin{tabular}{|c||c|c|}
\hline
& $\Mach_0=1.2$ & $\Mach_0=1.5$ \\
\hline
\begin{tabular}{c} \\ \hline NS \\ ENS \\ Diff. \end{tabular}
&
\begin{tabular}{c|c|c}
$\Delta\rho$, kg/m$^3$ & $\Delta u$, m/s & $\Delta T$, K \\
\hline
$8.164$ & $87.83$ & $56.73$ \\
$6.057$ & $69.3$ & $44.11$ \\
$26$\% & $21$\% & $22$\%
\end{tabular} &
\begin{tabular}{c|c|c}
$\Delta\rho$, kg/m$^3$ & $\Delta u$, m/s & $\Delta T$, K \\
\hline
$19.5$ & $198.5$ & $143.4$ \\
$16.43$ & $179$ & $126.6$ \\
$16$\% & $10$\% & $12$\%
\end{tabular}\\
\hline\hline
& $\Mach_0=2$ & $\Mach_0=3$ \\
\hline
\begin{tabular}{c} \\ \hline NS \\ ENS \\ Diff. \end{tabular}
&
\begin{tabular}{c|c|c}
$\Delta\rho$, kg/m$^3$ & $\Delta u$, m/s & $\Delta T$, K \\
\hline
$34.99$ & $356.4$ & $311.5$ \\
$30.41$ & $334.4$ & $284.5$ \\
$13$\% & $6$\% & $9$\%
\end{tabular} &
\begin{tabular}{c|c|c}
$\Delta\rho$, kg/m$^3$ & $\Delta u$, m/s & $\Delta T$, K \\
\hline
$54.26$ & $631.1$ & $765.9$ \\
$47.51$ & $602.6$ & $708.6$ \\
$12$\% & $5$\% & $7$\%
\end{tabular}\\
\hline
\end{tabular}
\caption{Magnitudes of shock transitions for all studied scenarios, as
  well as their relative differences in \%.}
\label{tab:magnitudes}
\end{table}

\section{Summary}

In the present work we study the effect of the Enskog correction on
the steady shock transitions in the supersonic flow of a hard sphere
gas, modeled via the Enskog--Navier--Stokes equations \cite{Abr17}.
First, we find that the Enskog-corrected speed of sound is somewhat
larger than the conventional speed of sound, and weakly depends on the
density of the gas in addition to its temperature. We then look for
one-dimensional solutions in the form of nondispersive waves, which
travel in a fixed direction at a constant speed. Equating the speed of
the reference frame to the speed of such a wave allows to reduce the
Enskog--Navier--Stokes equations into a system of two ordinary
differential equations for a spatial variable. Observing that a shock
solution should asymptotically approach two distinct uniform gas
states before and after the shock transition, we conclude that the
shock transitions are, in fact, the heteroclinic orbits which connect
distinct fixed points of this system. We then compute the fixed points
of the system exactly, and examine the heteroclinic orbits
numerically. Generally, we find that the Enskog correction affects the
shock transition in two different ways. First, the Enskog correction
leads to an overall weaker shock magnitude in comparison with the
conventional Navier--Stokes equations (that is, for the same pre-shock
state, the post-shock state has lower density and temperature, and
higher velocity).  Second, the Enskog correction increases the spatial
thickness of the shock transition, making the shock wave ``smoother''.
These effects are relatively more prominent at low Mach numbers, and
appear to be similar to what was observed in \cite{GreRee}.

\ack The author thanks Gregor Kova\v ci\v c for useful
discussions. The work was supported by the Office of Naval Research
grant N00014-15-1-2036, and by the Simons Foundation grant \#636144.

\end{document}